\documentclass[aps,pre,groupedaddress,amssymb,amsmath,pra,floatfix,showpacs]{revtex4}
\usepackage{epsfig}

\begin{document}
\title{Signal induced Symmetry Breaking in Noise Statistical Properties of Data Analysis}
\author{L. Perotti, D. Vrinceanu and D. Bessis}
\affiliation{Department of Physics, Texas Southern University, Houston, Texas\\
77004 USA}
\date{\today}

\begin{abstract}

From a time series whose data are embedded in heavy noise, we construct an Hilbert space operator ( J-operator) whose  discrete spectrum represents the signal while the essential spectrum located on the unit circle, is associated with the noise. Furthermore the essential spectrum, in the absence of signal, is built from roots of unity (``clock " distribution). These results are {\it independent} of the statistical properties of the noise that can be Gaussian, non-Gaussian, pink or even without second moment (Levy).

The presence of the signal has for effect to break the clock angular distribution of the essential spectrum on the unit circle. A discontinuity, proportional to the intensity of the signal, appears in the angular distribution.  The sensitivity of this method is definitely better than standard techniques.

We build an example that supports our claims.

%Pad\'{e} Approximations built on the time-series Z-transform.

\end{abstract}

\pacs{07.05.Kf, 07.05.Rm, 02.70.Hm}

\maketitle

\section{Introduction}

Analysis of highly ``noisy" time-series is crucial to progress in many fields, spanning from fundamental research such as gravitational wave detection, to practical applications such as e.g. Nuclear Magnetic Resonance Spectroscopy as applied to nuclear waste and/or brain and breast cancer detection, localization of underground oil fields, and early detection of structural faults in mechanical systems, just to mention a few.

``Noise" can encompass both uncorrelated (or slightly correlated) background and/or an abundance of unwanted signals.
While the latter ``Noise" can only be dealt with through some prior knowledge of the unwanted signal itself, and therefore requires methods tailored to each specific case considered, the existence of {\it universal} statistical properties for a wide class of noises, both white or slightly coloured (e.g. pink) \cite{noi} allows the design of general denoising techniques for it.

A well known tool in data analysis is the Z-transfom of the given time series, which is an extension to the complex plane of the Fourier transform. The Z-transforms of signal and noise have -in the limit of an infinite time series- completely different analytical properties, which we use to separate signal from noise. While the only singularities of the Z-transform of pure noise form a natural boundary on the unit circle, the singularities of the Z-transform of the signal component are isolated poles. We estimate the Z-transform of a time series by a suitable Pad\'{e} Approximant (a rational function whose expansion reproduces the Z-transform up to a given order) and analyze its poles and zeros. Two classes of poles and zeros clearly emerge: poles associated with the signal are isolated and (for weak noise) stable, i.e. their position in the complex plane does not change when increasing the length of the time sequence. 
Poles and zeros associated with the noise are instead randomly placed in the vicinity of the unit circle, organized in pole-zero pairs called ``Froissart doublets" \cite{dou1,dou2,dou3,dou4}, so as to mimic the natural boundary. Moreover, they  have statistical distributions in the complex plane which display universal properties \cite{noi}. Namely:

1) The radial distributions of both poles and zeros are Lorentzian, with widths depending only on the time-series length  $N$. For large $N$, this dependence goes approximately as $W\simeq \alpha \ln {(N)}/N$, the global multiplicative factor $\alpha$ being $0.404 \pm 0.005$ for the poles, and $0.52 \pm 0.01$ for the zeros.

2) The radial distributions of the poles is centered on 1; the radial distribution of the zeros is instead centered {\it on a point larger than 1}. The distance $D$ of this point from unity approaches zero when $N$ goes to infinity, again according to $D\simeq \alpha \ln {(N)}/N$, the global multiplicative factor $\alpha$ being $0.62 \pm 0.01$.

3) The phase distributions are uniform, approaching the $[N/2]$ roots of unity when $N$ goes to infinity (``clock" behaviour), {\it the deviation being Gaussian}.  

While the Z-transform is linear, the construction of a Pad\'{e} Approximant is clearly not so. For finite data series this results in a ``contamination" of signal and noise. Thus, the positions of the signal poles for different lengths of the time series and/or different realizations of the noise are distributed on an area that grows with the intensity of the noise itself.

This contamination works both ways; the question we have therefore posed ourselves is what are the effects of the signal on noise; in particular on the above statistical properties of poles and zeros, which we know to be universal. 

What we find is that, even for signal to noise ratios so small that signal poles cannot be effectively distinguished from the noise ones, the presence of a signal can be detected from its effect on the statistical distribution of the noise poles and zeros.

Injection of a small signal over a noise background modifies the distribution of poles and -even more markedly- of the zeros: increasing the signal amplitude, the center of the radial distribution of the zeros approaches the unit circle; This shift of the zeros is not uniform in angle but is more pronounced for zeros close to the the signal pole. Moreover, the deviation of the angular position of the zeros from the roots of unity stops being random and develops a step in correspondence of the position of the signal pole; consequently, the distribution of the deviation of the zeros from the roots of unity becomes non-Gaussian. 

As we shall see the presence of these effects allows detection of noisy signals with signal to noise ratio (SNR) down to $0.004$ , where we have defined $SNR =(P_s/P_n)$ where $P_s$ is the average signal power and $P_n$ is the average noise power.

\section{Summary of the method: the J-matrix formalism}

Starting point of our analysis is the Z-transform
\begin{equation}
Z(z)=\sum_{n\geq 0}s_{n}z^{-n}  \label{1}
\end{equation}
of a data series $s_{0},s_{1},s_{2},.....s_{n},.......$

Usually the choice of Pad\'{e} Approximant to the above Z-transform falls on the $[n-1/n]$ type, whose numerator and denominator are respectively polynomials of degree $(n-1)$ and $n$, or on the ``diagonal" $[n/n]$ one. 
This choice is justified by noting that for a finite ensemble of damped oscillators, the discretized data will
read
\begin{equation}
s_{k}=
\mathop{\displaystyle\sum}
\limits_{p}A_{p}e^{i\omega _{p}\frac{k}{N}T} \qquad
k=0,1,2,...,N-1\qquad \omega _{p}=2\pi f_{p}+i\alpha
_{p}\label{data}
\end{equation}
where $A_{p},f_{p},$ and $\alpha _{p}$ are the complex peak amplitude (including the initial phase), the frequency and the damping factor of the $p^{th}$ oscillator. $T$ is the recording time and $N$ the number of data.
The Z-transform therefore tends, when the number of data points goes to infinity, to
\begin{equation}
Z(z)=\mathop{\displaystyle\sum}\limits_{k=0}^{+\infty }s_{k}z^{-k}=
\mathop{\displaystyle\sum}
\limits_{k=0}^{+\infty }z^{-k}\mathop{\displaystyle\sum}\limits_{p}A_{p}e^{i\omega _{p}\frac{k}{N}T}=
\mathop{\displaystyle\sum}
\limits_{p}A_{p}
\mathop{\displaystyle\sum}
\limits_{k=0}^{+\infty }(z^{-1} e^{i\omega _{p}\frac{T}{N}})^{k}=\mathop{\displaystyle\sum}
\limits_{p}\frac{A_{p}}{1-z^{-1} e^{i\omega _{p}\frac{T}{N}}}
\end{equation}
which is clearly a $[n-1/n]$ rational function in $z^{-1}$, with $n$ equal to twice the number of oscillators.

For pure noise, instead, the organization of poles and zeros in Froissart doublets \cite{dou1,dou2,dou3,dou4} is best approximated by a $[n/n]$ rational function in $z^{-1}$.

The sum of the two is therefore a $[n/n]$ rational function; on the other hand, a $[n-1/n]$ Approximant gives the right number of poles and zeros. When the number of data points is large compared to the number of signal oscillators, the two choices give very similar results.  

To numerically calculate poles and zeros of the Pad\'{e} Approximants (either of the $[n-1/n]$ type, or of the $[n/n]$ type) of the Z-transform of a finite time series, we build directly from the time series two tridiagonal Hilbert space operators, called J-Matrices, one for the numerator and one for the denominator. The eigenvalues of these matrices provide zeros and poles, avoiding general polynomial root finding procedures, which are notoriously unstable. Details of our method can be found in Ref. \cite{noi}. Knowledge of the positions of all poles and zeros also gives us the residues for all poles and therefore the amplitudes and phases of the signal oscillations.

\section{Results and sensitivity of the method}

In order to clearly visualize the statistical properties of our ``J-Pad\'{e} Transform" (JPT) we shall consider multiple time sequences all with the same signal but with different noise. 
We shall explain at the end of the present letter how the JPT nonlinearity  allows us to obtain such multiple sequences when they are not readily available, so that it can be gainfully applied in cases when only a small number of sequences may be available; or even just a single one. 
The results we now present were obtained using $200$ time sequences of $300$ data points each, so that the $149/150$ JPT was used. 

For our first visualization, we associate to each pole (zero) a positive (negative) Gaussian distribution of suitable width and sum the values these distributions assume on the points of a lattice in the complex plane. Finally, we sum the distributions thus obtained for all the different data time sequences.

Poles $z_s$ associated to the signal are both {\it isolated} (i.e. not coupled to zeros) and {\it stable} (i.e they do not move when we change the number of data points used to calculate the JPT or when the noise is slightly modified, so that, apart from numerical and truncation errors, their position is the same for each time sequence); their contributions therefore sum up into peaks both close to the unit circle (constant amplitude signals) and away from it (damped signals). If the signal amplitude is not too small, these peaks are easily detectable .

The pole and the zero of each Froissart doublet are close to each other on the unit circle where they are most dense and their separation grows when we move away from the unit circle itself. 
On the other hand, noise poles and zeros in the tail of the respective distributions have highly random positions, and therefore their contributions tend to average to zero when summed over the different time sequences. Noise poles close to the unit circle are very close to the associated zeros, their contributions would therefore tend to average to zero for each time sequence, were not for the second of the statistical properties of noise listed above: the two radial distributions (of poles and zeros respectively) are slightly shifted. The net result in the complex plane is a circular ridge of radius slightly smaller than one surrounded by a circular trench of radius slightly larger than one, just like the walls and the moat of a medieval city.

%figures (increasing sensitivity): 

The first observable effect of the injection of a signal over the background noise is a shift of the center of the zeros radial distribution toward the unit circle: apart from some saturation effects when the signal is very large, the distance of the center of the zeros radial distribution from the unit circle is proportional to $K(N)^{-\rho}$, where $K(N)$ is a constant depending on the number $N$ of data points only, and $\rho$ is the magnitude of the signal residue. 

\begin{figure}[htbp]
\centering\epsfig{file=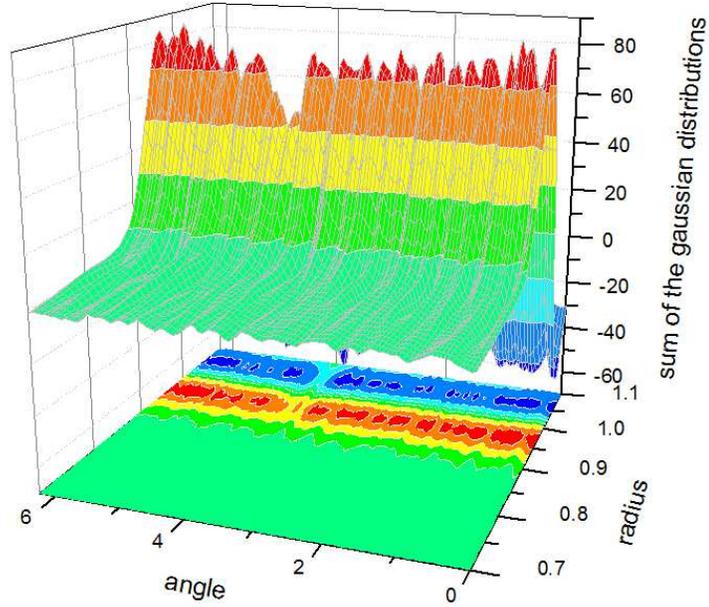,width=0.7\linewidth}
%\vspace{0.6in}
\caption{(Color online) Sum of the Gaussian distributions in polar coordinates. The signal is given by a single pole whose inverse is in $|z_s|=0.9$ (the signal amplitude halves in about $7$ data points) and $\varphi=4.0$, with residue $1.0$. The noise is complex Gaussian with unit amplitude; the seeds are changed for all $200$ time sequences of $300$ points. The gap in both ridge and trench at the angle of the signal pole is evident.}
\label{fig1}
\end{figure}

For weak signals, this radial shift is not uniform over the entire circle but is more pronounced close to the signal pole, thus creating a gap both in the ridge and the trench right at the angular position of the signal pole: see Fig. \ref{fig1}.

The discrete Fourier transform, being -in the complex plane- the restriction of the Z-transform to values of z located at the roots of unity, is only aware of what can be seen on the unit circle itself, approximately halfway between the ridge induced by the poles and the trench due to the zeros. It is therefore clear from Fig. \ref{fig1} that it is not able to detect this effect, as it looks where it is at its most weak.
 
\begin{figure}[htbp]
\centering\epsfig{file=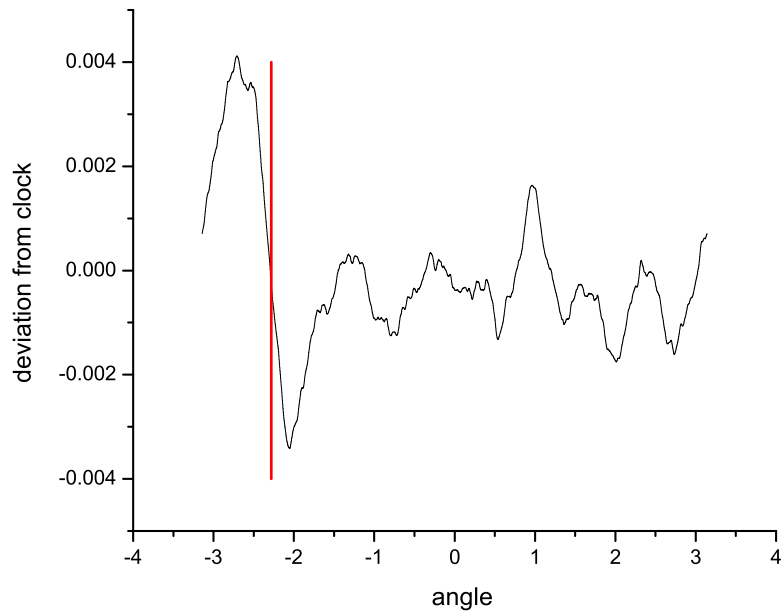,width=0.7\linewidth}
%\vspace{0.6in}
\caption{(Color online) Angular deviation of the position of the zeros of the JPT from the roots of unity (deviation from clock) Vs. angle, smoothed over 2000 points. The data points are the same as in Fig. \ref{fig1}, but the signal residue is reduced $0.5$. The thick red vertical line indicates the angular position of the signal pole, perfectly halfway through the downwards step of the deviation function.}
\label{fig2}
\end{figure}

Fig. \ref{fig1} suggests that there might also be an angular perturbation of the distribution of JPT poles and zeros. This is indeed the case. Fig \ref{fig2} shows the deviation from clock (i.e. from the roots of unity) of the JPT zeros for a signal amplitude half of that of Fig.  \ref{fig1}: a very clear step is evident at the angular position (frequency) of the signal. A similar step is displayed by the poles' distribution, only somehow smaller, probably due to the signal poles. The sign of the step depends on our definition of deviation. Defining the deviation from the n'th root of unity as $d_n = r_n - \varphi_n$, where $r_n$ is the angular position of the n'th root of unity and $\varphi_n$ is the angular position of the n'th zero (ordered in ascending angle order) then the step is always downwards.

\section{Generating the samples}

Can the effects we have seen above help us detect signals more effectively than Fourier transform analysis and methods derived from it, such as e.g. wavelet analysis?

We have clearly seen these effects, but only for a case when a high number of equivalent data series was available. Even if it is sometimes possible to gather such data, this is not usually the case. 

The next question is therefore how to get a sufficient number of data sequences capable of producing independent or only slightly correlated JPT's.  

If only a small number of data sequences is available, then the analysis can be applied to the ensemble of all (or part of) the combinations of the given data sequences: {\it since the Pad\'{e} Approximant is non linear, the resulting poles and zeros will not be linearly dependent}, even if some correlation remains. An added advantage is that while the signal amplitude sums as the number $m$ of samples in each combination, the noise only grows as $\sqrt{m}$, thus giving a $m$ increase in the SNR.

\begin{figure}[htbp]
\centering\epsfig{file=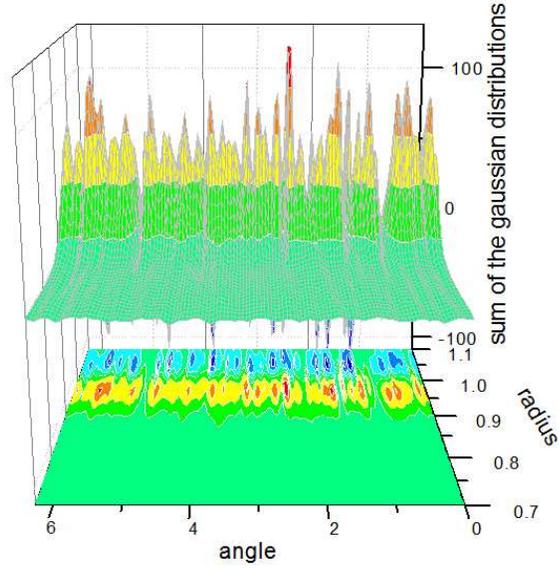,width=0.7\linewidth}
%\vspace{0.6in}
\caption{(Color online) Sum of the Gaussian distributions in polar coordinates. The signal is given by a single pole whose inverse is in $|z_s|=0.9$ and $\varphi=1.0$, with residue $0.5$. The noise is complex Gaussian with unit amplitude. The $163$ combinations of $4$ or more of the $8$ given data sequences are considered. Two gaps are evident: one at the signal angle, and the other at $\varphi \simeq 5.0$.}
\label{fig3}
\end{figure}

\begin{figure}[htbp]
\centering\epsfig{file=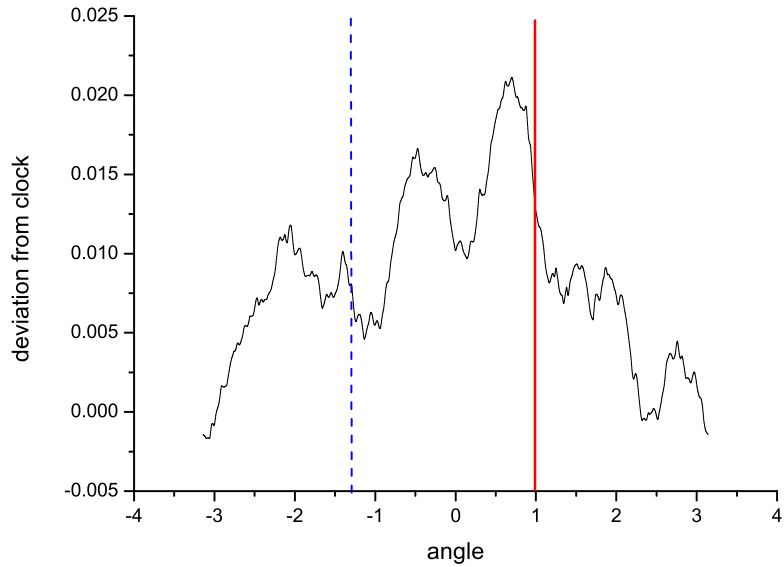,width=0.7\linewidth}
%\vspace{0.6in}
\caption{(Color online) Angular deviation of the position of the zeros of the JPT from the roots of unity (deviation from clock) Vs. angle, smoothed over 1800 points. The signal is the same as in Fig. \ref{fig3}. A clear step is visible at the signal angle, marked by a full red vertical line. At the position of the of the fake gap in Fig. \ref{fig3} (marked by a blue dashed line) no significant step is visible. }
\label{fig4}
\end{figure}

As an example, let's again consider a single signal pole, located in $|z_s|=0.9$ and $\varphi=1.0$ with residue (maximum signal amplitude) $\rho=0.5$ added to a Gaussian noise having unit standard deviation ($SNR=0.004$). We take only $8$ data sequences and consider all the $163$ combinations of $4$ or more of them. The results are shown in Figures \ref{fig3} and \ref{fig4}. The sum of the Gaussian distributions now shows two gaps in the ridge: one corresponding to the signal pole; and a fake one at $\varphi=5.0$. The zeros deviation from clock shows at least three sizeable down steps, the largest one right at $\varphi=1.0$, while none is at $\varphi=5.0$. Comparison of Figures \ref{fig3} and \ref{fig4} thus gives us the signal pole alone.

Using only the original $8$ sequences, we do get a comparable deviation from clock plot, but the sum of the Gaussian distributions shows no ridge gap corresponding to the step at $\varphi=1.0$, thus leaving us in doubt. 

\begin{figure}[htbp]
\centering\epsfig{file=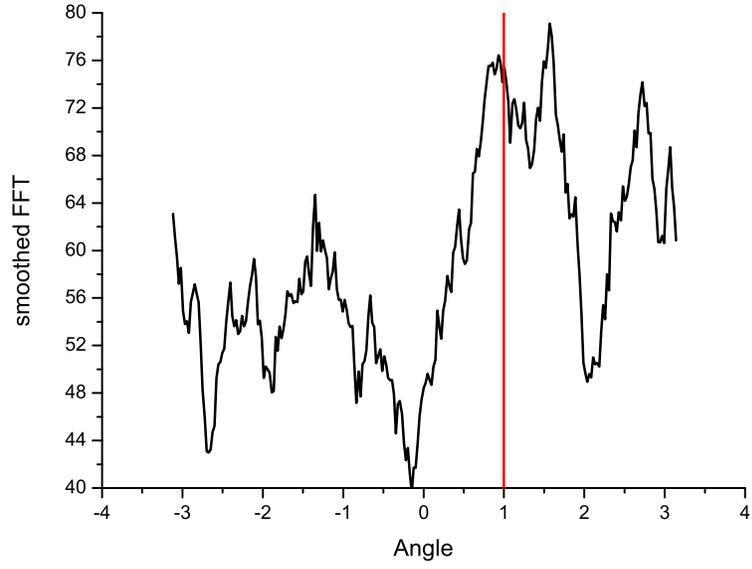,width=0.7\linewidth}
%\vspace{0.6in}
\caption{(Color online) Smoothed Fast Fourier Transform (FFT) for the data from Fig. \ref{fig3}. The vertical line marks the angular position of the signal pole: it's one of three peaks having the same height.}
\label{fig5}
\end{figure}

The Fourier transform being linear, the best we can do is to analyze the sum of all $8$ signals, so as to increase the SNR to about $0.03$. Fig. \ref{fig5} shows that, even with some smoothing of the transform, the signal peak is not recognizable among other comparable peaks.

Finally, what if we do have only one data sequence?

Most data sequences are often oversampled and undersampling is routinely used to decimate the data points and speed up calculations. Taking advantage of this, we can build from
the original time sequence $m$ different undersampled sub-sequences, all with the same time step (``interlaced sampling"): the first sequence comprising the data points $1,m+1,2m+1,...$, the second one comprising the data points $2,m+2,2m+2,...$, and so on. Undersampling is lossless for JPT \cite{noi}, unless the sampling rate is close to the expected signal frequency, in which case it would cause a modulus $2\pi$ uncertainty on the frequency.

We now have multiple time sequences all with the same signal (phase differences only affect the residues, but not the positions of the poles) but with different noise. 

If the number of undersampled sequences is too low, we can again use their linear combinations: our tests show results equivalent to those shown in Figures \ref{fig3} and \ref{fig4}, even if the phases of the signal differ from one sequence to another, so that there is in general no SNR gain in taking the combinations.

When the sampling rate is to low to allow any convenient undersampling (i.e. when the sampling rate is close to the expected signal frequency and undersampling would therefore cause a modulus $2\pi$ uncertainty on the frequency), summation can be performed over approximants of different order. A second -more attractive- possibility is to add different small amounts of noise to the same time sequence to again obtain a number of time sequences with the same signal and different input noise \cite{barone}. This latter procedure can be also used to improve statistics when some undersampling can be used. Note that its effectiveness again depends on the nonlinearity of our method; it would therefore be useless in an approach based on the Fourier transform.

\section{ Conclusions}

We have thus discovered that a significant increase in sensitivity can be achieved by looking at the perturbation of the statistical properties of the noise poles and zeros caused by the signal, rather than at the signal poles themselves.

\section{Acknowledgments}

We thank Professor Marcel Froissart, from College de France, for discussions and suggestions.

We thank Professor Carlos Handy, Head of the Physics Department at Texas Southern University, for his support.

Special thanks to Professor Mario Diaz, Director of the Center for Gravitational Waves at the University of Texas at Brownsville: without his constant support this work would have never been
possible.

Supported by sub-award CREST to the center for Gravitational Waves, Texas University at Brownsville, Texas USA

%\end{references}

%\vspace{0.9in}

%\newpage
%.

%\vspace{0.2in}

%\vspace{0.2in}

%\vspace{0.2in}

%\vspace{0.2in}

\end{document}